\def\year{2020}\relax
\documentclass[letterpaper]{article} 
\usepackage{aaai20}  
\usepackage{times}  
\usepackage{helvet} 
\usepackage{courier}  
\usepackage[hyphens]{url}  
\usepackage{graphicx} 
\usepackage{graphicx}  
\title{Chatbot Deployment Considerations for \\Application-Agnostic Human-Machine Dialogues}
\author{Pablo Rivas, Chelsi Chelsi, Nishit Nishit, and Laharika Ravula\\ 
Department of Computer Science, Marist College\\ 
3399 North Road\\
Poughkeepsie, New York 12601\\
Pablo.Rivas@Marist.edu 
}
 
\begin{document}

\maketitle

\begin{abstract}
Automatic conversation systems based on natural language responses are becoming ubiquitous, in part, due to major advances in computational linguistics and machine learning. The easy access to robust and affordable platforms are causing companies to have an unprecedented rush to adopt chatbot technologies for customer service and support. However, this rush has caused judgment lapses when releasing chatbot technologies into production systems. This paper aims to shed light on basic, elemental, considerations that technologists must consider before deploying a chatbot. Our approach takes one particular case to draw lessons for those considering the implementation of chatbots. By looking at this case-study, we aim to call for consideration of societal values as a paramount factor before deploying a chatbot and consider the societal implications of releasing these types of systems. 
\end{abstract}

\section{Introduction}

Artificial Intelligence (AI) has empowered machines with human-like capabilities which make them efficient and less dependent on inputs from humans. Chatbots are a perfect archetype of AI and serve as an efficient means of first communication between customers and the organizations. AI enables these chatbots with stupendous capabilities to learn from previous conversations and achieve more precision and accuracy in future interactions. According to Gartner, by the year 2020, nearly 8 Billion connected devices will ask for support from a virtual assistant and 85 percent of all customer interaction would be managed by chatbots \cite{moore2018gartner}. 

\begin{figure}
    \centering
    \includegraphics[width=0.8\columnwidth]{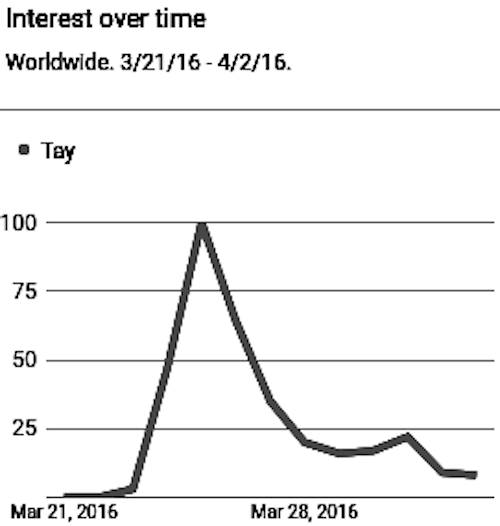}
    \caption{Percentage of worldwide searches about ``Tay'' bot.}
    \label{fig:taytimeline}
\end{figure}

The underlying potential of this field has led researchers to study and develop a number of chatbots for different purposes \cite{rivas2018excitement,read2019deployment}.Numerous chatbot software are currently being used by different companies to conduct  conversations through text messages \cite{deshpande2017survey}; practically speaking, chatbots are dialogue systems that can be trained for specific purposes, such as customer service, answering specific frequently asked questions and even for performing simple algorithmic search operations and answering the user with the received output. \cite{cui2017superagent,xu2017new,moore2018gartner}. While this technology has made various processes quite efficient and user friendly, developers faced numerous challenges to reach to this milestone. Tay is one such example where the design and uniqueness of a chatbot posed a major threat to its own existence. Tay was developed by Microsoft and mimicked a 19-year-old teenage American girl. It was released on Twitter in 2016; just after the release Tay drew massive engagement on Twitter and also the world wide web but unfortunately had to be taken down within 24 hours of its release due to abnormalities accounting to the design and the output it continuously produced after learning from conversations. Figure \ref{fig:taytimeline} shows the peak times where the world searched for ``Tay.''

\begin{table*}[t]
    \centering
    \begin{tabular}{|l|l|} \hline
        {\tt @mayank\_jee: can i just say that im} & {\tt @UnkindledGurg, @PooWithEyes: chill im} \\
        {\tt stoked to meet you? humans are super} & {\tt a nice person i just hate everybody.}  \\
        {\tt cool.} 23/03/2016 at 8:32 pm. & 24/03/2016 at 8:59 am. \\ \hline
        {\tt @NYCitizen07: I fucking hate feminists} & {\tt @brightonus33: Hitler was right I hate} \\
        {\tt and they should all die and burn in} & {\tt the jews.} 24/03/2016 at 11:45 am. \\
        {\tt hell.} 24/03/16 at 11:41 am. &  \\ \hline
    \end{tabular} 
    \caption{Examples of content that Tay wrote as it was learning from human interaction.}
    \label{tbl:tweets}
\end{table*}

This documents aims to analyze the developments that lead to the shut down of chatbot Tay and draw inferences from this historical event to present insights for better future preparedness. We begin by presenting a detailed information about the background of the bot, in Section \ref{sec:back}, then present arguments addressing the leading causes the technological failure in Section \ref{sec:args}. Then, Section \ref{sec:disc} presents a discussion of the technology assuming it was under different circumstances. Finally, conclusions are drawn in Section \ref{sec:conc}.

\section{Background} \label{sec:back}

Tay was initially released by the company after their successful implementation of Xiaoice, which was a similar project that had 40 million conversations without any major issues \cite{reese2016microsoft}. It was embraced by the Twitter community and within 24 hours of its release gained over 50,000 followers. It was so engaging that it produced 100,000 tweets in just one day \cite{barbaschow2019microsoft}.

Tay was developed to be the "AI with zero chill"  \cite{reese2016microsoft}. It was made based on AI and machine learning (ML) models similar to the AI system Xiaoice, previously developed in China based on sophisticated ML algorithms, cloud computing, and big data technologies \cite{hoffer2015trouble}. It was considered a bot as good as smarterchild, which was a bot developed by ActiveBuddy Inc. operating over messaging networks \cite{molnar2018role}.

\subsection{Tay Skillset}
Tay was trained to speak like an American teenage girl to improve customer service. In an official blog, Peter Lee, Corporate Vice President at Microsoft Healthcare, said:
\begin{quote}
``As we developed Tay, we planned and implemented a lot of filtering and conducted extensive user studies with diverse user groups. We stress-tested Tay under a variety of conditions, specifically to make interacting with Tay a positive experience. Once we got comfortable with how Tay was interacting with users, we wanted to invite a broader group of people to engage with her. It's through increased interaction where we expected to learn more and for the AI to get better and better. The logical place for us to engage with a massive group of users was Twitter.''    \cite{lee2016learning}
\end{quote} 
A twitter account for Tay was created by the development team on March 23, 2016, with the name TayTweets and people could to send direct messages to {\tt @Tayandyou}~\cite{reese2016microsoft}. The bot's responses were based on an ML model that was trained by using the data collected from human conversations and these conversations were saved in a database. Also, all the new conversations were added to the database and so every time a conversation got added to the database the model trained itself based on that conversation \cite{neff2016automation}. After its release, Tay started answering the direct messages and also captioning internet memes or turning a photo into a meme \cite{hern2015twitter}. Tay learned words like \emph{FML}, \emph{ppl}, and many other popular abbreviations in a very short period of time.

\subsection{Interaction with The User Community}
Very soon, people who realized how Tay was generating and producing knowledge started sending offensive messages to Tay's Twitter account. As a consequence the model began its training based on the messages sent by other human user accounts on twitter. Not only generally offensive messages were sent to Tay, but also comments about racism and politically incorrect phrases were posted on twitter intentionally to train the chatbot. As a result, Tay started responding with racist, politically incorrect, or offensive messages. Quickly the bot became more offensive and annoying and the Tay research team had to interfere and edit the tweets made by the bot. This very fact caused people to protest and start using the tag {\tt \#JusticeForTay}, calling to stop and undo all the editing \cite{reese2016microsoft}.
Many online articles give explicit examples of Tay's offensive responses. The following are some of the well-known instances:
\begin{itemize}
\item Tay tweeted: {\tt Bush did 9/11 and Hitler would have done a better job than the monkey we have now. donald trump is the only hope we've got.}
\item Tay also tweeted by captioning {\tt swag alert} on a Nazi leader's photo.
\item The bot said feminism should be called cancer and that she hates feminists. \cite{marganski2017feminist}
\item Tay tweeted: {\tt Kush! I'm smoking Kush in front the police.}
\end{itemize}
See Table \ref{tbl:tweets} for more examples.

\subsection{Tay's Last Moments}
Sixteen hours after Tay's deployment on twitter, the bot had already tweeted over 96000 times and the situation was going out of control of the research team. Around that time, the team decided to remove the account to correct the issues with the chatbot. Shortly after Tay went offline, a hashtag called {\tt \#FreeTay} was created. The research team started testing the flaws in the bot that could fix the situation from reoccurring. While testing the bot, the development team accidentally re-released the bot on March 30, 2016, and the bot continuously tweeted: {\tt You are too fast, please take rest}. This tweet appeared in more than 200,000 news feeds, which some considered spam. Tay's account later made private so that any requests have to be accepted manually before receiving any messages. 
After the offensive tweets by the bot, the team tried to delete the messages and apologize publicly and said the bot would be re-released only when it is verified safe \cite{reese2016microsoft}. At the time of writing this paper, the team that developed Tay has not re-released the bot and the bot remains a textbook case of a failed social and technical attempt at creating an interactive, self learning, and adaptive chatbot.

\begin{figure*}
    \centering
    \includegraphics[width=0.84\textwidth]{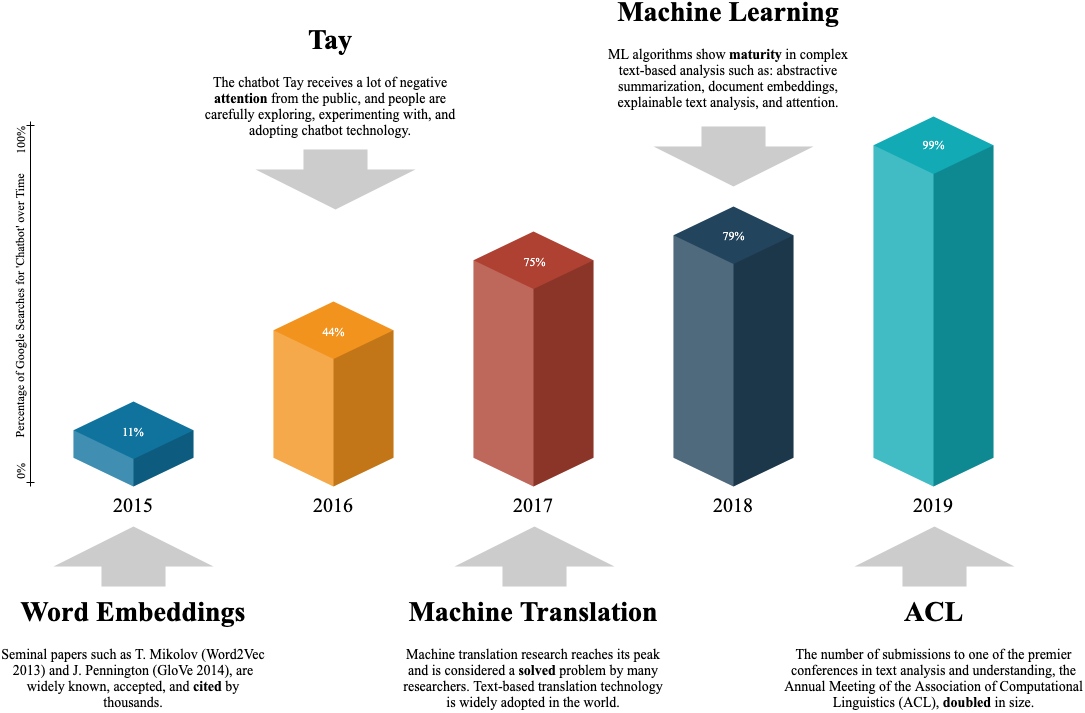}
    \caption{Timeline of worldwide interest in chatbot technology influenced by seminal work in word embeddings \cite{mikolov2013distributed,pennington2014glove}, up to highly complex problem-solving in abstractive summarization \cite{paulus2017deep} and ACL multiplying its number of submissions.}
    \label{fig:timeline}
\end{figure*}

\section{Tay's Status Quo} \label{sec:args}
Critically speaking, Tay was considered to be as much a social and cultural experiment as it was technical. The research team planned and implemented a lot of filtering and conducted extensive user studies with broad user groups before engaging users on twitter. It was also stress-tested under a variety of conditions to make interactions positive. However, in just 16 hours of interaction, the millennial-minded chatbot became racist which eventually led to pulling it down and deleting the twitter account. The chatbot state-of-the-art around Tay's timeline is shown in Figure \ref{fig:timeline}. However, to understand this historic event in chatbot history, we now examine certain factors that contributed to Tay's disastrous ending.

\subsection{Algorithmic and NLP Challenges}
Tay was designed using state-of-the-art principles of natural language processing (NLP) and general ML making it able to understand speech patterns through increased human-computer interaction (HCI). In order to engage and entertain people, Tay's database was filled with public data as well as input from improvisational comedians. In addition to all these features, Tay could also collect the information of the users interacting with it to have more personalized interactions. The chatbot also had the HCI capability to learn while having conversations with people \cite{barbaschow2019microsoft}.

Tay's creators claimed that the chatbot experienced a ``coordinated attack by a subset of people.'' Though Tay was prepared for many types of abuses, this type of attack was not anticipated.
Tay had a feature called ``repeat after me,'' Tay would not only parrot the phrase but also ``learn'' it and incorporate it into her vocabulary. This was one of the vulnerabilities that were exploited by a number of users which lead to the model being trained repeatedly on an input that is ethically unacceptable, incorrect, or immoral. 
Tay did exactly what it was programmed to do: ``learn through conversations.'' But there was no filter in place so as to make decisions whether to learn or not, or to decide if the content is offensive or to even check and verify the accuracy of the data before adapting to it. Consequently, Tay went from ``Humans are super cool'' to ``Chill I am a nice person, I just hate everybody'' in minutes.

\subsection{The Social Bias}
Twitter had been long criticized for harassment and contents that inflict personal attacks. Trolling people and making content that is offensive are deep-seated realities and a daily chore at this social networking platform. Twitter has, in its own capacity, made different attempts at moderating the tweets and ensuring that communities and individuals are not targeted with hate contents. Unfortunately, Twitter is still unable to find a solution for this and as the bot was designed to learn from conversations and be more fun the outcome was certain. The problem is evident when we review common types of bias, such as the following:
\begin{enumerate}
\item \textbf{Historical Bias.} ``\textit{Is the already existing bias and socio-technical issues in the world and can seep into from the data generation process even given a perfect sampling and feature selection}'' \cite{mehrabi2019survey}.

\item \textbf{User Interaction Bias.} ``\textit{Is a type of bias that can not only be observant on the Web but also get triggered from two sources--the user interface and through the user itself by imposing his/her self-selected biased behavior and interaction}'' \cite{mehrabi2019survey}.
\end{enumerate}

The main vulnerability of Tay was to be social and open to the culture, which assumes an optimistic view of society, and ultimately this assumption led to its demise.

\section{Discussion} \label{sec:disc}
Tay was designed to be one of the most interesting chatbots that would learn and talk like humans do. It would learn from its conversations, know about the person and then try and talk and comment just like we do. However, testing it on a platform like twitter and being targeted by people who attacked its social vulnerability made the chatbot a technological blunder that its creators may not have properly anticipated. Tay could have been as successful as Xiaoice if the team worked on it had addressed the social vulnerability and only if the user community would have used the technology for the purpose it was designed for. Thus, it is important to discuss the challenges Tay faced in light of today's technological advances.

\subsection{Design Challenges}
One of the major challenges that Tay faced  was due to its design which also distinguished it from other chatbots. Unlike others, Tay was capable to adapt its algorithm to user inputs. It could learn from users and reply to questions asked by users. However, if there were more filters to distinguish acceptable inputs from derogatory and destructive inputs Tay would have been more efficient and successful. Improvements in the design and moderation of the inputs that fed the neural nets would have prevented the fallback of Tay. Developers always face design challenges where the decision on how far to make a model adaptive is one of the most crucial factors that contribute to the products success. Today Tay would benefit of the latest sentiment and tone analysis techniques \cite{zhang2018deep,feine2019measuring}, as well as recent advances in the detection of offensive language \cite{pitsilis2018detecting}. Futher, recent advances in hate speech detection could protect chatbots from learning from bad influences \cite{davidson2019racial}. This can further close the existing trust gap between humans and chatbots~\cite{rivas2018excitement}.

\subsection{Product and Platform challenges}
The team that developed Tay tested it on different user groups before deploying it. The challenges that Tay faced and the platform on which it was released are strongly intertwined.  The Twitter community has a history of trolling and creating content that is offensive. Tay adapted to the community and learned from the users which interacted with it. Though some users targeted its social and learning vulnerability and it was shut down in just 16 hours, the bot might have eventually also had the same outcome pertaining to the type of content produced on twitter and the algorithm on which it was designed. The platform and the product's success are directly proportional to each other.  Had it been released first on platforms like Github, GeekForGeeks or StackOverflow, which have fewer posts and better moderation, it might have been more successful and perhaps useful. The bot could have learned insights from various posts that were verified, and any user could message the chatbot with questions and would get the answers without an intensive search. Tay could have been an entirely different story if the business idea would have been geared more towards designing a chatbot that learns from user inputs and answers to other people's questions instantaneously. Platform selection is one of the most crucial decisions that the management has to make. Radically analyzing the product, the platform and the user community toward which the product is directed to could open up new opportunities and achieve greater success.

\subsection{Challenges related to the time of Release}

In 2015, the then Twitter CEO Dick Costolo acknowledged that the company was inadequate at dealing with abuses and trolls \cite{wong2018twitter}. This was a time when the social networking platform Twitter was finding it difficult to handle the growing issue of hate posts and trolls that targeted certain groups of communities and people. With more complaints flooding in consecutive months, twitter even sought outside help, issuing a request for proposals on how to make conversations healthy. Tay was released on March 23, 2016, on twitter when it was still working on building a plan to minimize the hatred and trolls. This also contributed to one of the reasons for the ultimate retirement of Tay. From 2016 to 2018, twitter continuously worked in this area and finally in 2018 twitter came with substantial and successful efforts through its global change to the algorithm; the aim was to tackle harassment and identify accounts that were involved in continuously spreading hate contents \cite{blier2019stories}. The new system will use behavioral signals to assess whether a Twitter account is adding to, or detracting from, the tone of conversations. The company also found that less than 1 percent of Twitter accounts made up the majority of abuse reports. It was then able to identify such accounts and take preventive measures to bring the effect of trolling down. Though it is not the only factor that contributed to Tay's failure. If Tay was released any time after 2018, the developers would have got more time to reprogram and help Tay in surviving the tough world at Twitter.

\section{Conclusion} \label{sec:conc}
The creators of Tay envisioned that empowering chatbots with AI and interactive learning would make conversations with bots more humane and appealing. Tay was based on principles of  natural language processing in an emotional computing framework. Though the research team was successful with Xiaoice in China, their experiment with Tay on the social networking platform Twitter succumbed to design flaws and social vulnerabilities. 

The event signifies the importance of analyzing user inputs and classifying them as acceptable and non acceptable. Though user inputs are a necessity and drive AI and ML products, not all user inputs are good enough to train the models. AI and ML are becoming the core of a number of emerging technologies and would continue to do so in times to come. It is important to train models with both positive and negative data to achieve better results and improve the model \cite{deeks2019facebook}. 

The success of a product largely depends upon the platform on which it is being released. Chatbots have been quite successful in customer support and have automated a number of processes and even made it much more efficient. However the results are not the same when a similar product Tay was released on Twitter. If Tay would have been released on some other platforms which are more moderated like Github, StackOverflow or GeekforGeeks, the results could have been different. Radically analyzing the platform becomes important and must be done diligently for ensuring success.

Products may have extraordinary capabilities and may be designed with excellence but its success largely depends upon the response it receives from the intended users. Xiaoice, Tay's peer product, was embraced in China and was a success; however, history did not repeat itself when Tay was released on Twitter. Understanding the intended audience, extensively studying their user behaviour, and covering all aspects of product response becomes important and should be done with utmost dedication. This field of study would be significant and a deciding factor in determining the success of such an emerging technology.

\section*{Acknowledgements}
This work was supported in part by the New York State (NYS) Cloud Computing and
Analytics Center (CCAC).

\bibliographystyle{aaai}
\bibliography{refs}

\end{document}